\documentclass{elsart}

\usepackage{amssymb}
\usepackage{epsfig}

\def\F{\mathcal{F}}
\def\prc{Phys. Rev. C}
\def\pl{Phys. Lett.}

\begin{document}

\begin{frontmatter}

\title{Mass measurements and superallowed $\beta$ decay}

\author{J.C. Hardy}
\author{I.S. Towner}\footnote{Present address: 
Department of Physics,
Queen's University, Kingston, Ontario K7L 3N6, Canada}
\address{Cyclotron Institute, Texas A \& M University,                    
College Station, Texas  77843}
\author{G. Savard}
\address{Physics Division, Argonne National Laboratory, Argonne, Illinois 60439, USA
and Department of Physics, University of Chicago, Chicago, Illinois 60637, USA}

\begin{abstract}

A recent Penning-trap measurement of the masses of $^{46}$V and $^{46}$Ti leads to a $Q_{EC}$ value
that disagrees significantly with the previously accepted value, and destroys overall consistency
among the nine most precisely characterized $T=1$ superallowed $\beta$ emitters.  This raises the possibility
of a systematic discrepancy between Penning-trap measurements and the reaction-based measurements upon which
the $Q_{EC}$ values depended in the past.  We carefully re-analyze ($n$,$\gamma$) and ($p$,$\gamma$) reaction
measurements in the $24 \leq A \leq 28$ mass region, and compare the results to very precise Penning-trap
measurements of the stable nuclei $^{24}$Mg, $^{26}$Mg and $^{28}$Si.  We thus determine upper limits to
possible systematic effects in the reaction results, and go on to establish limits for the mass of
radioactive $^{26}$Al, to which future on-line Penning-trap measurements can be compared.  We stress the urgency
of identifying or ruling out possible systematic effects. 

\end{abstract}

\end{frontmatter}

\section{Introduction}
\label{s:intro}

Early in 2005, Hardy and Towner \cite{Ha05} published a complete survey
of all half-life, decay-energy and branching-ratio measurements
pertaining to 20 superallowed $0^+ \rightarrow 0^+$ decays.
For nine of these, the decay $ft$ values were determined with a
precision of 0.15\% or better.  These data, further adjusted for
radiative and isospin-symmetry-breaking  corrections \cite{To02},
yielded corrected $\F t$ values that were all the same to within
their derived uncertainties.  This is a highly satisfactory situation 
since it confirms the expectations of the Conserved Vector Curent (CVC)
hypothesis that the corrected $\F t$ values for Fermi transitions 
between states of a particular isospin should all be the same.
Indeed, this is a prerequisite if 
the up-down element, $V_{ud}$, of the Cabibbo-Kobayashi-Maskawa (CKM)
matrix is to be determined from the average $\F t$ value.  The singular
advantage that nuclei have over the neutron or the pion for
the determination of $V_{ud}$ is that there are many examples,
currently nine with good precision, that can be averaged
together to reduce the uncertainties while concurrently building 
confidence that no unforeseen pitfalls are lurking in the data. 

Later in 2005, the first Penning-trap mass measurement of one
of these nine most precisely determined superallowed transitions
was published \cite{Sa05}.  Of the nine transitions,
$^{46}$V had the largest uncertainty associated with its $Q$ value
and, for this reason, it was selected for the Penning trap measurement.
The result was startling.  The mass obtained for the parent, $^{46}$V, and
the daughter, $^{46}$Ti, both differed substantially from the
adopted values in the 2003 mass tables \cite{Au03}, and the mass difference
yielded a decay $Q$ value nearly three standard deviations away
from the average value quoted in the Hardy-Towner compilation \cite{Ha05}. 
 
\begin{figure}[t]
\vspace{-2.5cm}
\epsfig{file=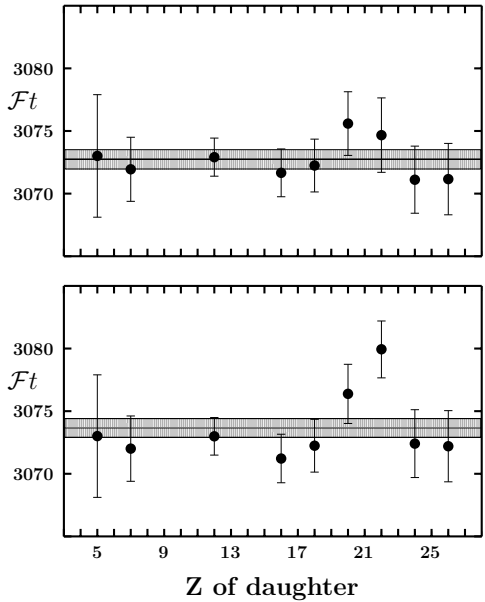,width=8.5cm}
\caption{Upper panel:  Corrected $\F t$ values from the Hardy-Towner
compilation \protect\cite{Ha05}.  Lower panel:  Revised $\F t$ values
after removing the $(^3$He$,t)$ measurements of Vonach {\it et al.} \protect\cite{Vo77}
and adding the Penning trap measurement of Savard {\it et al.} \protect\cite{Sa05}.}  
\label{fig1}
\end{figure}

All nine of the prime superallowed transitions feed stable daughter nuclei.  In these cases four 
principal methods have been used in the past to determine their decay $Q$ values:
{\it (a)} a $(p,n)$ threshold measurement, {\it (b)} a $(^3$He$,t)$
$Q$-value measurement, {\it (c)} a $Q$-value difference measurement (between two superallowed transitions)
with $(^3$He$,t)$ reactions on a composite target, and {\it (d)} a combination of
$(p,\gamma )$ and $(n,\gamma )$ measurements.  For $^{46}$V,
there had only been two previous measurements:  a threshold $(p,n)$
measurement of Squier {\it et al.} \cite{Sq76} giving $Q_{EC} = 7053.3(18)$ keV
and a $(^3$He$,t)$ measurement of Vonach {\it et al.} \cite{Vo77} giving
$Q_{EC} = 7050.41(60)$ keV.   The new Penning-trap result \cite{Sa05} of
7052.90(40) keV agrees with the former but is in strong disagreement
with the latter, which claims smaller error bars.  This latter $(^3$He$,t)$
result appeared in a publication \cite{Vo77} that included six other
$Q$ values in the prime series of superallowed emitters, all quoted with
similarly small uncertainties.  Since the energy calibration used nearly 30 years
ago in Ref. \cite{Vo77} cannot be reconstructed and most of its other six
results differ from more recent measurements, Savard {\it et al.}
\cite{Sa05} opted to remove all seven measurements
from the high-precision data set.  The consequences are dramatic.
In the upper panel of Fig.~\ref{fig1} we show the corrected
$\F t$ values as published in the Hardy-Towner compilation \cite{Ha05}
where all values are seen to be in accord.  In the lower panel, we
show the results of removing the $(^3$He$,t)$ measurements
of Vonach {\it et al.} \cite{Vo77} and including the Penning trap
measurement of Savard {\it et al.} \cite{Sa05}.  There is serious
deterioration.  The data are no longer all consistent as required
by the CVC hypothesis:  The $\F t$ value for $^{46}$V is significantly above the average.

From a purely experimental perspective, the new Penning-trap result for the $^{46}$V
$Q_{EC}$ value disagrees only with a single previous experiment \cite{Vo77}, one whose
results for other nuclei also appear to be aberrant.  Nevertheless, the $\F t$-value
anomaly it creates clearly raises suspicion and emphasizes the critical importance
of extending these Penning-trap measurements to other cases.  Suspicion is now further
aroused by a preliminary result \cite{Sa05b} for the mass of $^{42}$Sc, which may also
lead to a $Q_{EC}$ value that is higher than the previous average from reaction
measurements \cite{Ha05}.  At issue now is the question of whether $Q$ values measured
in Penning traps are consistently higher than those determined by other
methods, or whether $^{46}$V is an isolated occurrence.  If the
latter is the case, then the restoration of CVC consistency would
seem to depend on there being a deficiency in the current
nuclear-structure dependent corrections \cite{To02} obtained for $^{46}$V.
If the former is the case, then all $Q$-value measurements, past and
present, need to be scrutinized even more carefully for undetected
systematic effects.

In this paper, we critically examine two key reactions used
in the past to measure the $Q$ values of superallowed transitions. We
compare $(p,\gamma )$ and $(n,\gamma )$ reactions in the $24 \leq A \leq 28$
mass region with very precise (conventional) Penning-trap measurements of
stable nuclei, and set limits on possible undetected systematic effects
in these reactions.  We then derive a very conservative range for the
mass excess of radioactive $^{26}$Al based on these reactions alone.  When $^{26}$Al
is measured by (on-line) Penning trap, it should provide a critical test
for the presence of systematic effects.  If the Penning-trap result does not
lie within our range, then considerable urgency will have to be placed on
resolving the discrepancy and possibly completely re-evaluating all the
accepted superallowed $Q$ values.

\section{\label{1}A test of consistency}

The masses of $^{24}$Mg, $^{26}$Mg \cite{Be03} and $^{28}$Si \cite{Di94} have all been
measured with Penning-trap mass spectrometers.  These same masses can also be interconnected
by reaction $Q$ values, the masses of $^{24}$Mg and $^{26}$Mg being related by a pair of
($n$,$\gamma$) reactions, on $^{24}$Mg and $^{25}$Mg; and the masses of $^{26}$Mg and $^{28}$Si
being related by a pair of ($p$,$\gamma$) reactions, on $^{26}$Mg and $^{27}$Al.  This provides
a so-far unique situation in which Penning-trap and reaction measurements can be directly
compared at relatively high precision.  The reaction results are quoted to a precision of
80-300 eV and the Penning-trap results are quoted to 2-32 eV. 

\begin{table}
\caption{\label{table1}Mass excesses (in keV) of $^{24}$Mg and $^{26}$Mg as measured by Penning
trap; and the mass excesses of $^{25}$Mg and $^{26}$Mg as determined from the $^{24}$Mg mass
excess by addition of the measured $Q$ values for the $^{24}$Mg($n$,$\gamma$) and $^{25}$Mg($n$,$\gamma$)
reactions taken from various different measurements.}
\vspace{3mm}
\begin{tabular}{lclll}
\hline
\hline
\multicolumn{1}{l}{Method}
& \multicolumn{1}{c}{Ref.}
& \multicolumn{1}{c}{ $^{24}$Mg }
& \multicolumn{1}{c}{ $^{25}$Mg}
& \multicolumn{1}{c}{ $^{26}$Mg} \\[0.5mm]
 \hline
Penning trap & \cite{Be03} & $-13933.576(13)$ & & $-16214.529(32)$ \\
\\[-5mm]
($n$,$\gamma$)  & \cite{Is80} & & $-13192.760(120)$ & $-16214.340(170)$ \\
& \cite{Hu82} & & $-13193.090(140)$ & $-16213.680(460)$ \\
& \cite{Pr90} & & $-13192.900(80)$ & $-16214.680(120)$ \\
& \cite{Wa92} & & $-13192.910(50)$ & $-16214.770(70)$  \\
& \cite{Ch05} & & $-13192.790(40)$ & $-16214.630(50)$ \\
\hline
\hline
\end{tabular}
\end{table}

Since the trap and reaction techniques are very different, they should not share any common
sources of systematic error.  Therefore, by combining the two ($n$,$\gamma$) reaction $Q$ values
and comparing the result with the ($^{26}$Mg$-^{24}$Mg) mass difference obtained from the
Penning-trap measurements, we can have some measure of systematic errors associated with that
reaction; and similarly by combining the two ($p$,$\gamma$) reaction $Q$ values and comparing
the result with the ($^{28}$Si$-^{26}$Mg) mass difference we can determine possible systematic
errors in ($p$,$\gamma$) measurements.  The outcome will allow us to judge the validity of
reaction $Q$-value results in general, but particularly among short-lived exotic nuclei, for
which Penning-trap measurements are unavailable or are known with less precision.

\subsection{\label{a}($n$,$\gamma$) reactions }

\begin{figure}[t]
\epsfig{file=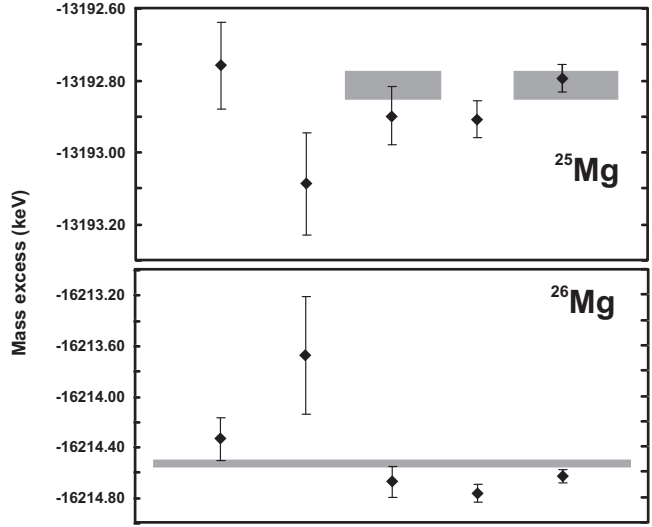,width=8.5cm}
\caption{The mass excesses of $^{25}$Mg and $^{26}$Mg as determined from the Penning-trap-measured
mass excess \cite{Be03} of $^{24}$Mg with the inclusion of the $^{24}$Mg($n$,$\gamma$) reaction
(for $^{25}$Mg) and the ($n$,$\gamma$) reactions on both $^{24}$Mg and $^{25}$Mg (for $^{26}$Mg).  In
each panel the five data points correspond to references \cite{Is80,Hu82,Pr90,Wa92,Ch05} (left
to right).  The shaded band in the bottom panel is the experimental mass excess of $^{26}$Mg
as determined by the Penning trap measurement \cite{Be03}.  The broken band in the top panel is the
average of only the two $^{25}$Mg data points \cite{Pr90,Ch05} through which it passes.} 
\label{fig2}
\end{figure}

In the first row of table \ref{table1} we give the mass excesses of $^{24}$Mg and $^{26}$Mg as
measured with a Penning trap \cite{Be03}.  There have been five independent measurements of
the $Q$ values for ($n$,$\gamma$) reactions on the three stable isotopes of magnesium
\cite{Is80,Hu82,Pr90,Wa92,Ch05}.  We have taken the $^{24}$Mg($n$,$\gamma$)$^{25}$Mg and
$^{25}$Mg($n$,$\gamma$)$^{26}$Mg $Q$ values as obtained in each measurement and combined them
with the $^{24}$Mg mass excess from the first row of the table to derive a mass excess for
$^{26}$Mg.  These five results appear in rows 2 through 6 of the last column, where they can each
be compared with the higher precision $^{26}$Mg mass excess from the top row.  This comparison
is also illustrated in the bottom panel of Fig \ref{fig2} where the ($n$,$\gamma$) results appear as
solid diamonds with error bars, and the Penning-trap result is presented as a shaded band.  It can
be seen that two of the ($n$,$\gamma$) data points for $^{26}$Mg differ significantly from the
Penning-trap value, one of them by three standard deviations.

\begin{table}
\caption{\label{table2}The $Q$ values for the $^{14}$N($n$,$\gamma$)$^{15}$N reaction as accepted at
various dates in the past 30 years.  The value $\Delta$ gives the difference between that $Q$ value
and the one accepted in 2003.}
\vspace{3mm}
\begin{tabular}{llcl}
\hline
\hline
\multicolumn{1}{c}{Date}
& \multicolumn{1}{c}{$Q$ value (keV)}
& \multicolumn{1}{c}{~ $\Delta$ (keV)}
& \multicolumn{1}{c}{Ref} \\ [0.5mm]
 \hline
2003 & 10833.2961(11) & 0 & \cite{Au03} \\
1993 & 10833.234(11) & $-0.062$ & \cite{Au93} \\
1990 & 10833.232(20) & $-0.064$ & \cite{Wa90} \\
1983 & 10833.302(12) & +0.006 & \cite{Co83} \\ 
1977 & 10833.361(58) & +0.065 & \cite{Wa77} \\
\hline
\hline
\end{tabular}
\end{table}

We have examined the original ($n$,$\gamma$) publications \cite{Is80,Hu82,Pr90,Wa92,Ch05} in detail
in order to determine, if possible, how the $\gamma$-ray energies used in their calibration have
changed in the interim since the experiments were originally analyzed.  Three of the references
\cite{Is80,Pr90,Wa92} state that their calibrations relied importantly on $\gamma$ rays observed from the
$^{14}$N($n$,$\gamma$)$^{15}$N reaction, some of whose energies depended on the $Q$ value contemporaneously
accepted for that reaction.  As can be seen in Table \ref{table2}, there have been significant changes
in the $^{14}$N($n$,$\gamma$)$^{15}$N $Q$ value over the period of time involved
\cite{Au03,Au93,Wa90,Co83,Wa77}.  Nevertheless, one of the three magnesium ($n$,$\gamma$) measurements
\cite{Pr90} occurred at a
time when that $Q$ value \cite{Co83} coincidentally agreed with its modern value.  It is noteworthy
that this measurement produced a $^{26}$Mg mass excess that agrees well with the Penning-trap value
(see Table \ref{table2}).  For the other two measurements \cite{Is80,Wa92} based in part on the $^{14}$N
$Q$ value, the fact that only some of their calibration $\gamma$ rays will have been affected by changes
in that $Q$ value makes it impossible for us to adjust with any confidence the original magnesium $Q$
values to correct for these changes.  In fact, any simple scaling of the derived $Q$ values by the ratio
of the contemporary $^{14}$N $Q$ value to the presently accepted value leads to even larger discrepancies
between these results and the Penning-trap value for $^{26}$Mg.

Of the two remaining measurements, one \cite{Hu82} is calibrated relative to the $\gamma$ rays from
$^{28}$Al and $^{36}$Cl, neither of which offer any better opportunities to update the
original results from 1982.  The other is the most recent \cite{Ch05} and is as yet unpublished; it
is calibrated against several sources of delayed and prompt $\gamma$ rays.  Though the actual energies
used are not specified, we presume that they are up to date and need no adjustment.  This result also
is within 100 eV of the Penning-trap value for $^{26}$Mg -- although still slightly outside the very
tight uncertainties quoted.

This leaves two ($n$,$\gamma$) results \cite{Pr90,Ch05} that remain valid by modern standards and, from
Table \ref{table1} and Fig. \ref{fig2}, we can see that both results lie slightly below the
Penning-trap result, by 150(130) and 100(60) eV respectively.  Keeping in mind that each quoted
($n$,$\gamma$) result actually represents the successive application of two such measurements -- on
$^{24}$Mg and $^{25}$Mg -- we can certainly conclude that any systematic differences between Penning-trap
and individual ($n$,$\gamma$) measurements must be less than 100 eV.  We may also suggest, if not
conclude, that uncertainties originally quoted on ($n$,$\gamma$) results may have been assigned with
some degree of optimism.

In general, if ($n$,$\gamma$) measurements are to be used in the determination of precise masses, then
it is essential that the calibration energies originally used be examined carefully.  If they differ
significantly from modern more-precise standards, then there is no simple way that the results can be
updated.  If they are to be used at all in their uncorrected form, the quoted uncertainties must be
increased by at least the amount of the change in the calibration $Q$ value and probably by more.  Finally,
it must be added that our conclusions here are drawn from careful measurements that were performed
under clean conditions; not all ($n$,$\gamma$) measurements can necessarily be presumed to have such small
systematic uncertainties.

\subsection{\label{b}($p$,$\gamma$) reactions }

The mass excesses of $^{26}$Mg and $^{28}$Si have both been measured with a Penning trap \cite{Be03,Di94}.
The result for the former was listed in Table \ref{table1}; the latter appears in the first
column of table \ref{table3}.  These two nuclei are also related by the $Q$ values 
for the reactions $^{26}$Mg($p$,$\gamma$)$^{27}$Al and $^{27}$Al($p$,$\gamma$)$^{28}$Si.  In contrast
though to the situation with the ($n$,$\gamma$) $Q$ values discussed in the previous section, there
has only been one measurement of each reaction $Q$ value -- references \cite{Ma78a,Ma78b}, respectively --
published within the past 40 years.  The result of our combining the published $Q$ values from these
two measurements with the mass excess of $^{26}$Mg to obtain the $^{28}$Si mass excess is given in the
second column of the table, which is labeled ``uncorrected".  This result differs from the Penning trap
value by 710 eV, slightly more than the combined uncertainties.

\begin{table}[t]
\caption{\label{table3}Mass excess (in keV) of $^{28}$Si as measured by a
Penning trap, and as determined from the $^{26}$Mg mass (see Table \ref{table1}) by addition of the
measured $Q$ values for the $^{26}$Mg($p$,$\gamma$) and $^{27}$Al($p$,$\gamma$) reactions.
The process used to correct the ($p$,$\gamma$) result is described in the text.}
\vspace{3mm}
\begin{tabular}{lll}
\hline
\hline
\multicolumn{1}{c}{Penning trap \cite{Di94}}
& \multicolumn{2}{c}{($p$,$\gamma$) \cite{Ma78a,Ma78b}} \\[0.5mm]
\cline{2-3}
& \multicolumn{1}{c}{uncorrected}
& \multicolumn{1}{c}{corrected} \\[0.5mm]
 \hline
\\[-5mm]
$-21492.7968(19)$ & $-21492.090(640)$ & $-21492.590(360)$ \\
\hline
\hline
\end{tabular}
\end{table}

Since both these ($p$,$\gamma$) measurements took place in 1978, we again examined the original publications
in detail to establish what calibrations were used and how they have changed in the intervening 27 years. 
For the ($p$,$\gamma$) $Q$-value measurements it turned out to be possible to up-date the original results
with reasonable confidence and, in fact, to reduce the uncertainties originally quoted.  We deal first with
the proton resonance energies and then with the energies of the emitted $\gamma$ rays.

In both measurements, the proton resonance energies were calibrated relative to a particular
$^{27}$Al($p$,$\gamma$)$^{28}$Si resonance that had been established in 1970 to be at $E_p = 991.880(40)$ keV
by an absolute velocity technique \cite{Ro70}.  This value has been revised since, with
Wapstra \cite{Wa03} in 2003 recommending the value 991.830(50) keV although his recommendation apparently
did not include provision for the very precise remeasurement, 991.724(21) keV, made by Brindhaban and Barker
\cite{Br94} ten years earlier.  We choose a conservative approach and adopt the value 991.780(60) keV, which
encompasses both these recent values.  This amounts to a 100-eV shift in the proton-energy standard from the
time the ($p$,$\gamma$) experiments \cite{Ma78a,Ma78b} were calibrated.  When converted to the center-of-mass
system, this change increases the $Q$ values quoted in both experiments by 96 eV, and marginally increases their
uncertainties.

The basis for $\gamma$-ray energy calibration was also common to both experiments and depended on the energies of
$\gamma$ rays from the decay of $^{66}$Ga as given by Heath in 1974 \cite{He74}.  These $^{66}$Ga $\gamma$-ray
energies have been reviewed again quite recently by Helmer and van der Leun \cite{He00}: their recommended energies are
systematically shifted from the Heath values -- though generally not outside of Heath's quoted error bars -- and
their uncertainties are significantly smaller.

The $^{26}$Mg($p$,$\gamma$)$^{27}$Al $Q$ value measured in Ref. \cite{Ma78a} utilized two pairs of $\gamma$ rays in
$^{27}$Al, all of energy $\sim$4.5 MeV.  A comparison of the modern $^{66}$Ga $\gamma$-ray energies with the Heath
catalog values used in the original experiment indicates that the energies in this region have increased by approximately
70 eV.  The reaction $Q$ value originally quoted in the paper must therefore be increased by 140 eV to account
for this calibration shift.  At the same time, the quoted uncertainty, which was dominated by Heath's uncertainties
in the calibration energies, can be reduced significantly.  Incorporating the shifts both in the resonance proton
energy and in the calibration $\gamma$ rays, we revise the $^{26}$Mg($p$,$\gamma$)$^{27}$Al $Q$ value, which was
originally quoted \cite{Ma78a} as 8271.0(5), to 8271.2(3) keV.

The paper reporting the $^{27}$Al($p$,$\gamma$)$^{28}$Si $Q$ value \cite{Ma78b} specifically correlated the $\gamma$
rays in $^{28}$Si that were used to measure the $Q$ value with the $\gamma$ rays from $^{66}$Ga that were used to
calibrate them.  We could then correct each one for the known shift between the $^{66}$Ga energies originally
used \cite{He74} and their modern counterparts \cite{He00}.  This has the effect of increasing the originally
quoted $Q$ value by 220 eV from this cause alone and, once again, decreases its uncertainties.  Incorporating
the shifts both in the resonance proton energy and in the calibration $\gamma$ rays, we revise the
$^{27}$Al($p$,$\gamma$)$^{28}$Si $Q$ value, which was originally quoted \cite{Ma78b} as 11584.5(4), to 11584.8(2) keV.

Combining the corrected values just obtained for the two ($p$,$\gamma$) $Q$ values with the Penning-trap result for
the mass excess of $^{26}$Mg, we obtain the ``corrected" result for $^{28}$Si shown in the third column of Table
\ref{table3}.  It agrees well with the Penning-trap value, differing by only 210 eV, well under its 360-eV
uncertainty.  On this basis, and again remembering that two ($p$,$\gamma$) measurements were applied to obtain the
$^{28}$Si mass, we can conclude that any systematic differences between Penning-trap and ($p$,$\gamma$) results
must be less than 200 eV.  As with the ($n$,$\gamma$) measurements, here too reliable calibration has proved all
important but, at least in the ($p$,$\gamma$) experiments we considered, the original calibration methods were transparent
enough that corrections could be applied to adjust the results to modern standards.  Once again, though, we must
caution that not all ($p$,$\gamma$) measurements can necessarily be presumed to have such small systematic uncertainties.

\subsection{\label{c}Test outcome}

Our object in this consistency test was to search for possible systematic differences between modern, highly precise
Penning-trap measurements of nuclear masses and the results of reaction measurements that have been frequently used
in the past to determine mass differences.  Until recently, such reaction $Q$-value measurements were the only
precise way to obtain the masses of radioactive nuclei and, in cases where precision really mattered, uncertainties
as low as 120 eV have been quoted in published measurements.  In the recent survey of superallowed $\beta$ decay
\cite{Ha05}, the $Q_{EC}$ values for the nine most precisely known transitions all depended on such reaction $Q$
values.  Any discovery of pervasive systematic effects could have a significant impact on the weak-interaction
tests that have been based on superallowed data.

From the test cases we have studied, we can conclude that, if the calibration standards used in the original
measurements can be established and if they are consistent with modern values for those standards, then ($n$,$\gamma$)
$Q$ values can be considered potentially reliable down to uncertainties of, say, 100 eV.  For ($p$,$\gamma$) $Q$
values, the equivalent limit should be around 200 eV.  If the original energy-calibration standards are found to
have changed significantly, then the two reactions should be dealt with very differently: on the one hand, the
($n$,$\gamma$) $Q$ values cannot easily be updated and they should either be discarded entirely or have their
uncertainties increased by at least the amount of the calibration change; on the other hand, the ($p$,$\gamma$)
$Q$ values can be updated reliably if the changes in their calibration energies can be clearly documented.

\section{\label{2}Evaluation of the $^{26}$Al mass}

With guidelines now determined for the use of ($n$,$\gamma$) and ($p$,$\gamma$) $Q$-value measurements, we can
turn to establishing the most reliable reaction-based value for the $^{26}$Al mass excess.  This is an important
issue since the $Q_{EC}$ for the superallowed $\beta$ transition between $^{26}$Al and $^{26}$Mg is a key component
of stringent weak-interaction tests \cite{Ha05} and, as yet, its mass has not been measured with a Penning
trap.  Because of the tight restraints we have just been able to place on the reaction measurements in this
mass region, $^{26}$Al should also become a valuable test of Penning-trap measurements on radioactive
species.  The three Penning-trap measurements we considered in section \ref{1} were all of stable nuclei,
$^{24}$Mg, $^{26}$Mg and $^{28}$Si, and were conducted under ideal conditions: their uncertainties were
only a few eV.  Short-lived radioactive nuclei demand a more complex experimental arrangement on-line to
an accelerator, and furthermore the collected ions can only be retained in the measurement trap for a
relatively short time, thus limiting precision.  To date, such measurements quote uncertainties of a few hundred
eV and, in fact, none has yet been confronted by results of comparable precision based on a different technique.
Of course, it is also true that most reaction measurements have not been confronted by such an independent check
at this level of precision either.

Our first step is to determine the mass excess of $^{25}$Mg.  In Table \ref{table1} we show the result of
taking the $Q$ value for the $^{24}$Mg($n$,$\gamma$)$^{25}$Mg reaction from Refs \cite{Is80,Hu82,Pr90,Wa92,Ch05}
and combining them with the Penning-trap mass excess for $^{24}$Mg.  These values are also plotted in the
top panel of Fig. \ref{fig2}, where they show considerable scatter.  However, in section \ref{a} we
determined that only two sets of ($n$,$\gamma$) measurements \cite{Pr90,Ch05} were consistent with modern calibration
standards and yielded results for the $^{26}$Mg mass that were within 150 eV of the Penning-trap result.
Consequently, we accept only the values for the $^{25}$Mg mass excess associated with those two references and
average the results to obtain the value $-13192.812(37)$ keV.  This average is shown as a (broken) shaded
band in Fig. \ref{fig2}. 

There is a further concern.  We have noted already that our results for the $^{26}$Mg mass showed that the
($n$,$\gamma$)-based results were 100-150 eV lower (more negative) than the Penning-trap value.  Although, this
effect was only barely significant statistically, to be safe we choose to increase our derived $^{25}$Mg mass
excess by 60 eV and to increase its overall uncertainty to 100 eV to incorporate provision for possible
systematic effects.  The resultant final value of $-13192.752(100)$ keV also appears in
Table \ref{table4}.  Within uncertainties, it agrees with the value $-13192.830(30)$ keV listed in the 2003
Mass Tables \cite{Au03}.

\begin{table}
\caption{\label{table4}Input data for our determination of the mass excess of $^{26}$Al.}
\vspace{3mm}
\begin{tabular}{lr}
\hline
\hline
\multicolumn{1}{c}{Description}
& \multicolumn{1}{c}{Data (keV)} \\[0.5mm]
\hline
& \\[-5mm]
{\bf $^{25}$Mg mass excess from ($n$,$\gamma$):} & \\
~~~~~average, published values \cite{Pr90,Ch05} & $-13192.812(37)$~~ \\
~~~~~adjusted for possible systematics & $-13192.752(100)$ \\
& \\[-5mm]
{\bf $^{25}$Mg($p$,$\gamma$) $Q$ value:} & \\
~~~~~as published \cite{Ki91} & $6306.400(60)$~~ \\
~~~~~corrected for calibration changes & $6306.426(98)$~~ \\
~~~~~adjusted for possible systematics & $6306.426(200)$ \\
& \\[-5mm]
{\bf $^{26}$Al mass excess:} & \\
~~~~~not adjusted for possible systematics  & $-12210.27(11)$~~~ \\
~~~~~adjusted for possible systematics & $-12210.21(22)$~~~ \\
\hline
\hline
\end{tabular}
\end{table}

The next step is to examine measurements of the $^{25}$Mg($p$,$\gamma$)$^{26}$Al $Q$ value, from which
the $^{26}$Al mass can be derived.  Although there is only one publication \cite{Ki91} that claims precision
below 500 eV, it originates from the same laboratory as, and with one author in common with, the two
($p$,$\gamma$) papers \cite{Ma78a,Ma78b} we considered in section \ref{b}.  Even though it was published 13
years later, equivalent calibration adjustments to those made successfully on the earlier results might
be expected to achieve similar success with the later one.  The uncorrected $Q$ value appears as it was
published in Table \ref{table4}, and we now deal with what corrections must be applied so that it can be
made to conform with modern standards.  

The proton resonance energies in the $^{25}$Mg($p$,$\gamma$)$^{26}$Al measurement \cite{Ki91} were calibrated
relative to the same resonance in the $^{27}$Al($p$,$\gamma$)$^{28}$Si reaction as were the earlier measurements.
However, by the time of the later measurement, the resonance energy was taken to be 991.86(3) keV, 20 eV lower
than had been accepted previously.  As explained in section \ref{b}, more recent evaluations and measurements
have led us to adopt the value 991.780(60) keV for the resonance energy and, after conversion to the
center-of-mass system, this requires an increase of 76(60) eV in the total $Q$ value.

The excitation energies in $^{26}$Al that were used in Ref. \cite{Ki91} were taken from earlier work at the
same laboratory \cite{En88}, which measured the energies of $\gamma$ rays in $^{26}$Al based on those from
$^{66}$Ga decay.  Unfortunately, the $^{66}$Ga $\gamma$-ray energies were not listed in Ref. \cite{En88} but
instead were attributed to Alderliesten {\it et al.} \cite{Al93}, whose results did not actually appear in
print until 5 years later.  Assuming that the energies of the $^{66}$Ga $\gamma$ rays did not change in the
time between their use in calibration and their publication, we can compare the Alderliesten {\it et al.}
energies with the recent review by Helmer and van der Leun \cite{He00}: in the important region around 4 MeV
they are, on average, 35 eV higher than the currently accepted values but by a slightly larger amount at
higher energies.  Because the specific $\gamma$ rays used in calibration are impossible to ascertain at the
end of this tortuous path, we choose to reduce the $Q$ value by 50(50) eV, a conservative approach.

Applying these two corrections, which tend to cancel one another, we obtain the corrected value for the
$^{25}$Mg($p$,$\gamma$)$^{26}$Al $Q$ value of 6306.426(98) keV, where the uncertainty simply includes the
original experimental uncertainty and the uncertainties we have assigned to the calibration corrections.  In
section \ref{b}, we concluded that we could not rule out systematic uncertainties in ($p$,$\gamma$) $Q$-value
measurements below the level of 200 eV; so, once again to be safe, we increase the overall uncertainty on
our final value to 200 eV to accommodate possible undetected systematic effects.  Both our
calibration-corrected and final results for the $Q$ value are also listed in Table \ref{table4}.

Finally, we combine our values for the $^{25}$Mg mass excess and the $^{25}$Mg($p$,$\gamma$)$^{26}$Al
$Q$ value to obtain a value for the $^{26}$Al mass excess.  We do this first with the results that are not adjusted
for possible systematic effects: {\it i.e.} we use the first number listed for the $^{25}$Mg mass excess in
Table \ref{table4} and the second number listed for the ($p$,$\gamma$) $Q$ value.  The result is $-12210.27(11)$ keV.
Next, using the numbers that take account of possible systematic effects, we obtain the result $-12210.21(22)$ keV.
Both results compare favorably with the value $-12210.31(6)$ keV listed in the 2003 Mass Tables \cite{Au03} but, because
we incorporate updated calibration standards and, for our second result, include provisions for possible
systematic effects, our uncertainties are considerably larger.

One note of caution should be added.  Although the mass of $^{25}$Mg is derived from two independent but concordant
($n$,$\gamma$) measurements that also correctly obtain the mass excess for $^{26}$Mg, the link between $^{25}$Mg and
$^{26}$Al depends on a single ($p$,$\gamma$) measurement and must be regarded as less secure.  Although that
measurement came from the same laboratory, and used the same calibrations as ($p$,$\gamma$) measurements that
we have demonstrated to agree with precise Penning-trap results on stable nuclei, no corroborating ($p$,$\gamma$)
measurements on $^{25}$Mg exist.  Obviously, we cannot rule out an experimental aberration.

\section{\label{3}Conclusions}

We have carefully analysed measurements of ($n$,$\gamma$) and ($p$,$\gamma$) $Q$ values in the $24 \leq A \leq 28$
mass region and, after updating the results (where possible) to modern calibration standards, compared them with
very precise Penning-trap mass measurements of stable isotopes.  This has allowed us to set upper limits on
possible systematic effects of 100 eV for ($n$,$\gamma$) reactions and of 200 eV for ($p$,$\gamma$) reactions.  Based
on ($n$,$\gamma$) and ($p$,$\gamma$) reactions, we then established two values for the mass excess of
radioactive $^{26}$Al, one excluding and the other including the adjustment and limits we obtained for possible
unobserved systematic contributions.

Neither result should be regarded as a value we recommend for use in determining the superallowed
transition energy from $^{26}$Al.  Instead, together they provide a critical standard for reaction-based results,
to which a future on-line Penning-trap mass measurement can be compared.  If the Penning-trap result
turns out to lie within the limits of our first value (the one uncorrected for possible systematic effects), then
one can be reasonably confident that actual systematic effects are below the upper limits we set; in that case Penning-trap
measurements, when they proliferate, can simply be averaged in with the earlier reaction-based results.  If the
Penning-trap result lies outside the limits of our first value but inside the limits of our second value (adjusted for
systematics), then one must suspect that reaction measurements in general may suffer from undiagnosed systematic
effects; wherever their quoted uncertainties are in the few-hundred-eV region, they will need to be increased
accordingly.

If the Penning-trap result lies outside the range of even our systematics-adjusted result, then that could be a
sign of serious systematic difficulties, which could call into question all reaction-based measurements of superallowed
transition energies or, conversely, could cast doubt on the precision of on-line Penning-trap measurements of
radioactive isotopes.  This would require serious and urgent attention, particularly in the evaluation of superallowed
$\beta$ decay and its associated weak-interaction tests.

On-line Penning-trap measurements of $^{26}$Al (and $^{25}$Mg) to perform the comparison test are strongly
recommended.

This work was supported by the U.S. Department of Energy under Grant No. DE-FG03-93ER40773 and by the Robert A. Welch
Foundation under Grant No. A-1397.



\end{document}